\documentstyle[12pt,epsfig,rotating,a4]{article}
\begin{document}
%=======================================================================
\begin{center}
{\large\bf SIMUB -- a Monte Carlo Generator \\ 
           for Physics Simulation of $B$ Decays
\\
}
\vspace*{3mm}
{\bf A.A.Bel'kov, S.G.Shulga}\\[6pt]
{ Particle Physics Laboratory, JINR, \\
Francisk Skaryna Gomel State University }\\
\end{center}
\vspace*{3mm}
{\it Talk at the VIth International School-Seminar
     ``Actual Problems of High Energy Physics''
     August 7-16, 2001, Gomel, Belarus}

%=======================================================================
\begin{abstract}
   We present the SIMUB package developed at Dubna for MC generation of 
$B$ meson production and decays. 
   The starting version of the package includes lepton modes of $B$ 
decays, in particular, semileptonic decays $B\to D^{(*)}l\nu$ and 
``golden'' mode $B\to J/\psi(\to\mu\mu)\,\phi(\to KK)$ with taking into 
account all theoretical refinements including $B^0-\overline{B}^0$ 
oscillations and angular correlations.
\end{abstract}
%=======================================================================

   SIMUB is a Monte Carlo (MC) generator of $B$-meson production and 
decays which is under developing at Dubna for the Compact Muon Solenoid 
(CMS) Project at CERN (see SIMUB documentation in \cite{SIMUB}).
   The main motivation for this activity was that already existing 
generators do not take into account the theoretical refinements which are
of great importance for MC studies of $B$-decay dynamics.
   In particular, in the generators PYTHIA~\cite{PYTHIA}, QQ~\cite{QQ}, 
and EvtGen~\cite{EvtGen}, the time-dependent spin-angular correlations 
between the final-state particles are not included in the proper way for 
the so called ``golden'' decay 
$B^0_s(t),\overline{B}^0_s(t)\to J/\Psi(\to \mu^+\mu^-)\,\phi(\to K^+K^-)$.
   The dynamics of this decay is described by four-dimension probability 
distribution function depending on decay time and three physical angles.
   The algorithms of multidimensional random number generation have been
elaborated and then implemented in the package SIMUB to provide tools for 
MC simulation of sequential two-body decays 
$B^0(t)\,,\overline{B}^0(t)\to a(\to a_1 a_2)\,b(\to b_1 b_2)$ in accordance
with theoretical time-dependent angular distributions.

   This paper is organized as follows. 
   First, some general theoretical aspects are considered in the context of 
$B^0-\overline{B}^0$ oscilations and angular correlations.
   Then, we discuss the algorithms of multidimensional random number 
generation restricting ourself by detail consideration of only 
$B\to V_1(\to \mu^+\mu^-)\, V_2(\to P^+P^-)$ channel and its 
$B^0_s\to J/\psi\phi$ ``golden'' mode.
   Finally, we present a general information about SIMUB package and compare
time and angular distributions for ``golden'' decay mode generated by SIMUB
with those obtained by using PYTHIA. 

\vspace*{4mm}
%=========================================================================
{\large\bf 1. $B^0-\overline{B}^0$ mixing and time evolution of neutral\\
\hspace*{14mm}$B$-meson states}
%=========================================================================
\vspace*{2mm}

  The time dependence of $B^0\to f$ decays, in which $f$ is not a CP 
eigenstate, is not purely exponent due to the presence of 
$B^0-\overline{B}^0$ mixing.
  This mixing arises due to either mass difference or decay-width 
difference between the mass eigenstates of the $B^0-\overline{B}^0$ system.
  The time evolution of the state $|B^0(t)\rangle$ 
($|\overline{B}^0(t)\rangle$) of initially, i.e. at time $t=0$, present
$B^0$ ($\overline{B}^0$) meson can be described as follows:
\begin{eqnarray*}
&&|B^0(t)\rangle =  g_+(t)|B^0\rangle 
                  + g_-(t)|\overline{B}^0\rangle\,,\qquad
g_+(t=0) = 1\,,\quad g_-(t=0) = 0\,;
\nonumber \\ 
&&|\overline{B}^0(t)\rangle = \bar{g}_+(t)|B^0\rangle 
                             +\bar{g}_-(t)|\overline{B}^0\rangle\,,\qquad
\bar{g}_+(t=0) = 0\,,\quad \bar{g}_-(t=0) = 1\,.
\end{eqnarray*}

   The strong eigenstates $|B^0(t)\rangle$ and $|\overline{B}^0(t)\rangle$
are not eigenstates of the full Hamiltonian due to the weak interaction.
   Thus, they most generally evolve according to the Schroedinger equation:
\begin{eqnarray*}
i\frac{\partial}{\partial t}
\left(
\begin{array}{c}
|B^0(t)\rangle\\
|\overline{B}^0(t)\rangle
\end{array}
\right)
= \bigg({\bf M} - \frac{i}{2}{\bf \Gamma}\bigg)\left(
\begin{array}{c}
|B^0(t)\rangle\\
|\overline{B}^0(t)\rangle
\end{array}
\right)\,,
\end{eqnarray*}
\vspace{-7mm}
{\bf
\begin{eqnarray*}
{\bf M} =
\left(
\begin{array}{cc}
M_{11} & M_{12} \\
M_{21} & M_{22}
\end{array}
\right)\,,\quad
{\bf \Gamma} =
\left(
\begin{array}{cc}
\Gamma_{11} & \Gamma_{12} \\
\Gamma_{21} & \Gamma_{22}
\end{array}
\right)\,.
\end{eqnarray*}
}
   Diagonalization of full Hamiltonian ${\bf M}-(i/2){\bf \Gamma}$
(see \cite{richman} for more detail) gives
\vspace*{-4mm}
\begin{eqnarray*}
&&g_+(t) = \frac{1}{2}\Big( e^{-i\mu_+t} +  e^{-i\mu_-t}\Big)\,,\quad
  g_-(t) = \frac{\alpha}{2}\Big( e^{-i\mu_+t} -  e^{-i\mu_-t}\Big)\,;
\nonumber\\
&&\bar{g}_+(t) = g_-(t)/\alpha^2\,,\qquad\qquad\quad~
  \bar{g}_-(t) = g_+(t)\,.
\label{g_functions}
\end{eqnarray*}
   Here $\mu_{\pm} \equiv M_{L/H} - (i/2)\Gamma_{L/H}$ are eigenvalues 
of the full Hamiltonian corresponding to the masses and total widths of 
physical ``{\it light}'' and ``{\it heavy}'' eigenstates of full 
Hamiltonian $|B_{L/H}\rangle \equiv |B_\pm\rangle$, and
\begin{equation}
\alpha = \sqrt{\frac{M^*_{12}-\Gamma^*_{12}}{M_{12}-\Gamma_{12}}}
       \approx \sqrt{\frac{M^*_{12}}{M_{12}}}
       = \eta^B_{CP}
\label{eta_CP}
\end{equation}
is a phase factor defining CP transformation of flavor eigenstates of
neutral $B$-meson system: 
$CP|B^0(t)\rangle =\eta^B_{CP}|\overline{B}^0(t)\rangle$.

   In Eq.~(\ref{eta_CP}) we have made a good approximation
$\Gamma_{12}\ll M_{12}$ valid for $B^0-\overline{B}^0$ system.
   This approximation implies that there is no $CP$ violation in 
$B^0-\overline{B}^0$ mixing, i.e. the probability for $B^0$ to oscillate
to a $\overline{B}^0$ is equal to the probability of a $\overline{B}^0$ to
oscillate to a $B^0$:
$$
\mbox{Prob}(\overline{B}^0\to B^0)\approx\mbox{Prob}(B^0\to\overline{B}^0).
$$
   Such an asymmetry in mixing, which results from $|\alpha| \not= 1$,
is often reffered to as {\it indirect} CP violation.
   The situation with indirect $CP$-violation in $B^0-\overline{B}^0$ mixing 
is in contrast to the $K^0-\overline{K}^0$ system, in which
$$
\mbox{Prob}(\overline{K}^0\to K^0)>\mbox{Prob}(K^0\to\overline{K}^0).
$$
   The indirect $CP$-violation is the main source 
of $CP$-violation in neutral $K$ decays.

   From $CPT$ theorem it follows that
\begin{eqnarray*}
\nonumber
&&M_{11} = M_{22} \equiv M\,,\qquad 
  M_{12} = M_{21}^*\,;
\\
&&\Gamma_{11} = \Gamma_{22} \equiv\Gamma\,,\qquad\quad~
  \Gamma_{12} = \Gamma_{21}^*\,,
\end{eqnarray*}
where $M$ and $\Gamma$ are "mean" values: $M=(M_L+M_H)/2$, 
$\Gamma=(\Gamma_L+\Gamma_H)/2$. 
   The values $\Delta M=(M_H-M_L)$ and $\Delta\Gamma=(\Gamma_H-\Gamma_L)$
are used also.

   The explicit expressions for mixing probabilities as a function of time 
are given by
\begin{eqnarray}
\mbox{Prob}( B^0~\mbox{at}~t| B^0~\mbox{at}~t=0) & = &
\mbox{Prob}( \overline{B}^0~\mbox{at}~t| \overline{B}^0~\mbox{at}~t=0) 
\nonumber \\
& = &      \frac{1}{4}\Big[ e^{-\Gamma_L t} + e^{-\Gamma_H t}
                           + 2e^{-\Gamma t}\mbox{cos}(\Delta Mt)\Big]\,,
\nonumber \\
\mbox{Prob}( \overline{B}^0~\mbox{at}~t|B^0~\mbox{at}~t=0) & = &
\frac{1}{4}|\alpha|^2 \Big[ e^{-\Gamma_L t} + e^{-\Gamma_H t}
                           - 2e^{-\Gamma t}\mbox{cos}(\Delta Mt)\Big]\,,
\nonumber \\
\mbox{Prob}( B^0~\mbox{at}~t|\overline{B}^0~\mbox{at}~t=0) & = &
\frac{1}{4|\alpha|^2} \Big[ e^{-\Gamma_L t} + e^{-\Gamma_H t}
                           - 2e^{-\Gamma t}\mbox{cos}(\Delta Mt)\Big]\,,
\nonumber \\
\label{mix-prob}
\end{eqnarray}

   For case $|\alpha|=1$ (no indirect CP violation) in approximation 
$\Gamma_L \approx \Gamma_H$, the latter two expressions in 
Eqs.~(\ref{mix-prob}) can be rewritten in form
\begin{eqnarray*}
\mbox{Prob}( \overline{B}^0~\mbox{at}~t|B^0~\mbox{at}~t=0) =
\mbox{Prob}( B^0~\mbox{at}~t|\overline{B}^0~\mbox{at}~t=0)
\approx \mbox{sin}^2\left(\frac{\Delta Mt}{2}\right)
\end{eqnarray*}
   which coincides with the formula for $B^0-\overline{B}^0$ mixing 
probability used in PYTHIA~\cite{PYTHIA}:
\begin{equation}
{\cal P}_{flip} =
\mbox{sin}^2\left(\frac{xt}{2\langle\tau\rangle}\right)\,.
\label{ProbPYTHIA}
\end{equation}
   Here $\langle\tau\rangle = 1/\Gamma$ is the mean lifetime, and $x$ is
the mixing parameter \cite{PDG}:
\begin{eqnarray*}
x = x_d = \Delta M/\Gamma = 0.74\pm 0.02&
\mbox{in}& B^0_d -\overline{B}^0_d~\mbox{system;}\\
x = x_s = \Delta M/\Gamma > 19 &
\mbox{in}& B^0_s -\overline{B}^0_s~\mbox{system.}
\end{eqnarray*}
   In PYTHIA, the initial $B^0$ meson is allowed with the probability 
(\ref{ProbPYTHIA}) to decay like a $\overline{B}^0$ and vice versa.

\vspace*{4mm}
%=================================================
{\large\bf 2. Time evolution of decay amplitudes}
%=================================================
\vspace*{2mm}

   Time evolution of the amplitudes of transitions
$|B^0(t)\rangle ,\overline{B}^0(t)\rangle \to |f\rangle$ induced
by Hamiltonian $H_{eff}$ is represented by
\begin{eqnarray}
&&A_f(t)\equiv \langle f|{\cal H}_{eff}|B^0(t)\rangle
        =   g_+(t)\langle f|{\cal H}_{eff}|B^0\rangle
          + g_-(t)\langle f|{\cal H}_{eff}|\overline{B}^0\rangle \,,\qquad
\nonumber \\
&&\bar{A}_f(t)\equiv \langle f|{\cal H}_{eff}|\overline{B}^0(t)\rangle
        =   \bar{g}_+(t)\langle f|{\cal H}_{eff}|B^0\rangle
          + \bar{g}_-(t)\langle f|{\cal H}_{eff}|\overline{B}^0\rangle\,.
\qquad
\label{time_evol_transit1}
\end{eqnarray}
   If $|f\rangle$ is the {\it eigenstate of $CP$-operator}
\begin{eqnarray*}
  CP|f\rangle = \eta_{CP}^f|f\rangle\,,\quad (\eta_{CP}^f = \pm 1)\,,
\label{CP_f}
\end{eqnarray*}
the relations (\ref{time_evol_transit1}) can be rewritten in form
\begin{eqnarray*}
&&A_f(t) = \langle f|{\cal H}_{eff}|B^0\rangle
           \bigg[ g_+(t) 
                 +g_-(t)\frac{1}{\eta_{CP}^f\eta_{CP}^B}\xi_f\bigg]\,,
\\
&&\bar{A}_f(t) = \langle f|{\cal H}_{eff}|B^0\rangle
                 \bigg[ \bar{g}_+(t)
                       +\bar{g}_-(t)\frac{1}{\eta_{CP}^f\eta_{CP}^B}\xi_f
                 \bigg]\,.
\label{time_evol_transit2}
\end{eqnarray*}
   Here
\begin{eqnarray*}
\xi_f \equiv \frac{\langle f|{\cal H}_{eff}|\bar{B}^0\rangle }
                  {\langle f|{\cal H}_{eff}^{CP}|\bar{B}^0\rangle }
       = |\xi_f| e^{-i\phi_f}\,,
\end{eqnarray*}
where ${\cal H}_{eff}^{CP}\equiv CP{\cal H}_{eff}CP$, and $\phi_f$ is the
{\it CP-violating weak phase} \cite{fleischer1}.

   The CP-violating weak phase $\phi_f$ is introduced through interference
effects between $B^0-\overline{B}^0$ mixing and decay process with final
state $|f\rangle$ being the eigenstate of $CP$ operator.
   In case of decays $B^0_q,\overline{B}^0_q\to f$, where $q\in{d,s}$, the
value of weak phase $\xi^{(q)}_f$ can be expressed in terms of matrix
elements of the combinations ${\cal Q}^{jq}$ of four-quark operators and
Wilson coefficients involved in the low energy effective Hamiltonian
$H_{eff}$ (see Refs.~\cite{fleischer1,fleischer2}):
\begin{eqnarray}
\xi^{(q)}_f =
e^{-i\phi_{mix}^q}
\frac{\sum_{j=u,c} \lambda_j^{(q)}
      \langle f|{\cal Q}^{jq}|\overline{B}_q^0\rangle}
     {\sum_{j=u,c} \lambda_j^{(q)*}
      \langle f|{\cal Q}^{jq}|\overline{B}_q^0\rangle}\,.
\label{phase}
\end{eqnarray}
Here $\phi_{mix}^{q}$ is the $B_q^0-\bar{B}_q^0$ {\it mixing phase}:
\vspace{-2mm}
\[
\phi_{mix}^{q} \equiv 2\mbox{arg}(V_{tq}^*V_{tb})
           = \left\{ \begin{array}{ll}
                     ~~2\beta        & \mbox{for $q=d$}\,,\\
                     -2\delta\gamma & \mbox{for $q=s$}\,,\\
                     \end{array}
\right. \]
where $V_{ij}$ is the Cabibbo-Kobayashi-Maskawa (CKM) matrix elements while
$\beta$ and $\delta\gamma$ are standard parameters related with angles of the 
unitary triangles.

   In general, the observable $\xi_f^{(q)}$ suffers from large theoretical
uncertainties in the hadronic matrix elements 
$\langle f|{\cal Q}^{jq}|\overline{B}_q^0\rangle$.
   However, if the decays $B^0_q,\overline{B}^0_q\to f$ are dominated by a 
single CKM amplitude, the corresponding matrix elements in 
Eq.~(\ref{phase}) are canceled, and $\xi_f^{(q)}$ takes the simple form:
\begin{eqnarray*}
\xi_f^{(q)}  = e^{-i\phi_j^{(q)}}\,,\quad
\phi_j^{(q)} =  2[\mbox{arg}(V_{tq}^*V_{tb})-\mbox{arg}(V_{jq}^*V_{jb})]
             = \phi_{mix}^{q} - \phi^{j}_{dec}
\end{eqnarray*}
where  $q\in d,s$, $j\in u,c$, and $\phi^{j}_{dec}$ is a $CP$-violating
weak decay phase:
\[
\phi_{dec}^{j} \equiv 2\mbox{arg}(V_{jq}^*V_{jb})
= \left\{ \begin{array}{ll}
   -2\gamma & \mbox{for $j=u$ -- dominant $b\to u\bar{u}q$ }\,,\\
    ~~~~0   & \mbox{for $j=c$ -- dominant $b\to c\bar{c}q$ }\,,\\
          \end{array}
  \right.
\]
where $\gamma$ is the angle of unitary triangle.

\vspace*{4mm}
%====================================================================
{\large\bf 3. Transversity amplitudes and time-dependent\\
\hspace*{14mm}observables for decays $B(t)\to V_1V_2$}
%====================================================================
\vspace*{2mm}

   The decay $B^0(t)\to V_1V_2$, where $V_1$ and $V_2$ are vector mesons
in final state, is described in terms of three transverse amplitudes $A_0$,
$A_{||}$ and $A_\bot$ \cite{dighe1}:
\vspace{-4mm}
\begin{equation}
  A(B^0(t)\to V_1V_2) =   A_0(t)\frac{m_{V_2}}{E_{V_2}}
                            \epsilon^{*L}_{V_1}\epsilon^{*L}_{V_2}
                        - A_{||}(t)\frac{1}{\sqrt{2}}
                            \vec{\epsilon}^{\,*T}_{V_1}
                            \vec{\epsilon}^{\,*T}_{V_2}
                        - A_\bot(t)\frac{i}{\sqrt{2}}
                            \vec{\epsilon}^{\,*}_{V_1}
                            \vec{\epsilon}^{\,*}_{V_2}\hat{\vec{p}}\,.
\label{transv_ampl}
\end{equation}
   Here $E_{V_2}$ is the energy of the $V_2$ in the $V_1$ rest frame;
$\hat{\vec{p}}$ is a unit vector in the direction of the momentum of $V_2$
in $V_1$ rest frame; $\vec{\epsilon}_{V_1}$ and $\vec{\epsilon}_{V_2}$ are
the polarization three-vectors in the $V_2$ rest frame; 
$\epsilon^{L}\equiv \hat{\vec{p}}\cdot\vec{\epsilon}$ is the longitudinal 
component and $\vec{\epsilon}^{\,T}$ is the transverse component of the 
polarization vector.
   The decay $\overline{B}^0(t)\to V_1V_2$ is described in analogous way 
in terms of the transversity amplitudes $\bar{A}_0$, $\bar{A}_{||}$ and 
$\bar{A}_\bot$.

   Eq.~(\ref{transv_ampl}) corresponds to decomposition of final state of 
$B^0(t)\to V_1V_2$ decay over transversity basis which can be represented
in form of 3-dimensional vector
$$
  |f\rangle= (|(V_1V_2)_0\rangle ,|(V_1V_2)_{||}\rangle ,
              |(V_1V_2)_\bot\rangle)\,.
$$
   The transversity components are eigenstates of $CP$-operator:
\vspace*{-2mm}
$$
  CP|(V_1V_2)_f\rangle = \eta_{CP}^f |(V_1V_2)_f\rangle\,,\quad 
  (f = 0,||,\bot)
\vspace*{-4mm}
$$
with eigenvalues
\vspace*{-2mm}
$$
 \eta_{CP}^0 =1\,,\quad \eta_{CP}^{||} =1\,,\quad \eta_{CP}^\bot = -1\,.  
\vspace*{-2mm}
$$
    Thus, the Eq.~(\ref{time_evol_transit2}) with 
$\langle f|{\cal H}_{eff}|B^0\rangle \equiv A_f(0)$ can be applied to 
calculate the time evolution of transversity amplitudes $A_f(t)$ 
($f = 0,||,\bot$).

   Matrix element squared involves into consideration the following six 
time-dependent bilinear combinations of transversity amplitudes 
({\it observables}):
\vspace*{-2mm}
\begin{eqnarray}
&& {\cal O}_1(t)=|A_0(t)|^2\,,\quad {\cal O}_2(t)=|A_{||}(t)|^2\,,\quad 
   {\cal O}_3(t)=|A_\bot(t)|^2\,,
\nonumber \\
&& {\cal O}_4(t)=\mbox{Im}\Big( A_{||}^*(t) A_\bot(t)\Big)\,,\quad 
   {\cal O}_5(t)=\mbox{Re}\Big( A_0^*(t) A_{||}(t)\Big)\,,\quad
\nonumber \\
&&    {\cal O}_6(t)=\mbox{Im}\Big(A_0^*(t)A_\bot(t)\Big)\,,
\vspace*{-2mm}
\label{observ}
\end{eqnarray}
in case of decays $B^0(t)\to V_1 V_2$, and similar combinations of
$\bar{A}_f(t)$ ($f = 0,||,\bot$) for decays $\overline{B}^0(t)\to V_1 V_2$.

\vspace*{4mm}
%====================================================================
{\large\bf 4. The time-dependent angular distributions}
%====================================================================
\vspace*{2mm}

\begin{figure}[hbt]
\begin{center} 
\vspace*{-6mm}
\includegraphics[width=0.62\textwidth]{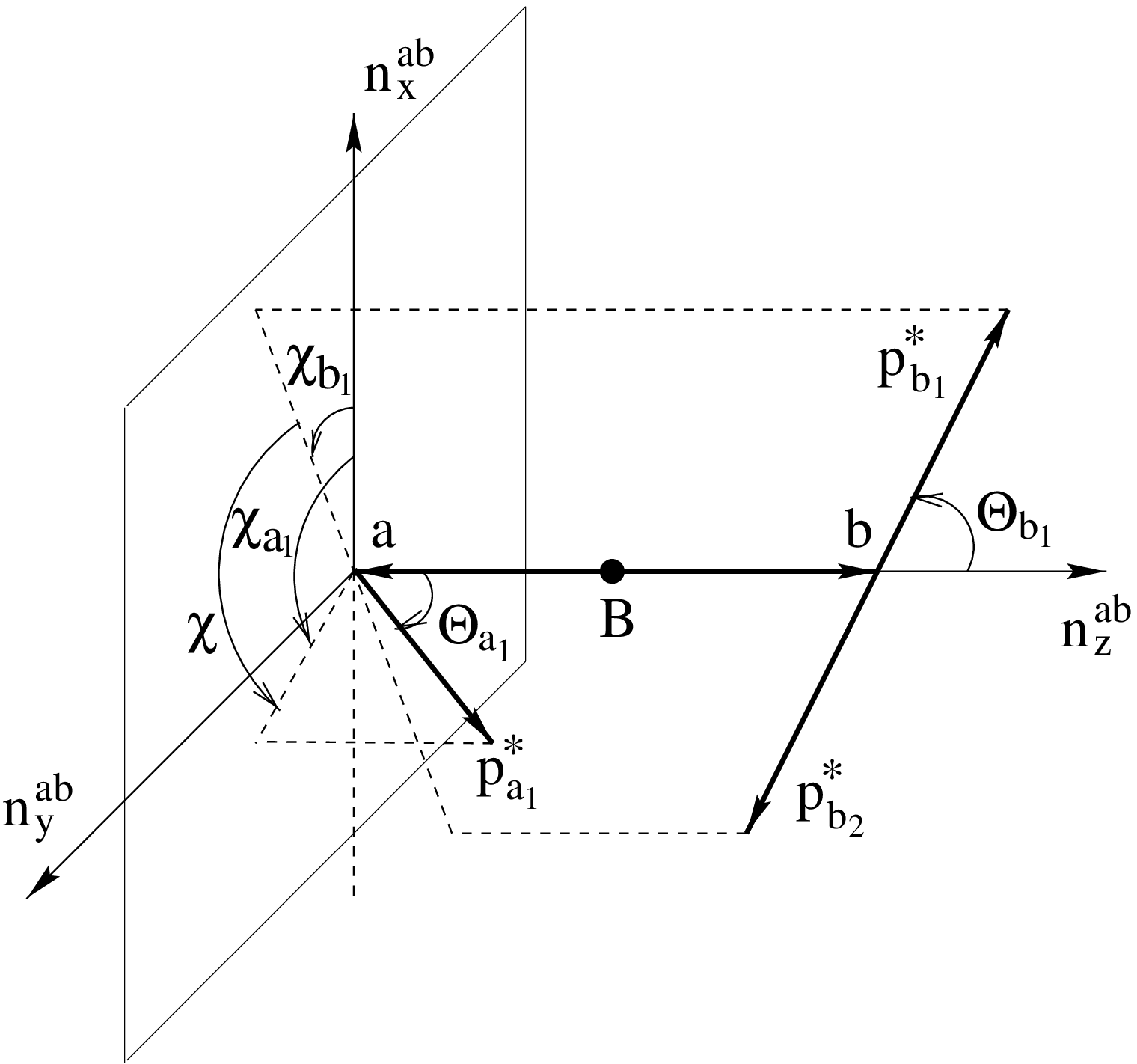}
\end{center}
\vspace*{-5mm}
\caption{\small The decay $B \to a(\to a_1a_2)\,b(\to b_1b_2)$ in 
                 helicity frame}
\label{B_dec_ANGL}
\end{figure}

   Let us consider sequential two-body decays 
\vspace*{-2mm}
$$
B^0(t)\,,\overline{B}^0(t)\to a(\to a_1 a_2)\,b(\to b_1 b_2)\,,
\vspace*{-2mm}
$$
 where $a$
and $b$ are vector mesons decaying into pairs of particles $a_1$, $a_2$ 
and $b_1$, $b_2$, respectively.
   The angular distributions for these decays are governed by spin-angular 
correlations (see \cite{jacob}-\cite{kutschke}) and involve three physical 
angles.
   In case of the so-called helicity frame \cite{kramer} these angles are 
defined as shown in Fig.~(\ref{B_dec_ANGL}).
   The $z$-axis is defined to be the direction of $b$-particle in the 
rest frame of the $B^0$. 
   The $x$-axis is defined as any arbitrary fixed direction in the plane 
normal to the $z$-axis. 
   The $y$-axis is then fixed uniquely via $y=x\times z$.
   The angles ($\Theta_{a_1}$, $\chi_{a_1}$) specify the direction of the 
$a_1$ in the $a$ rest frame while ($\Theta_{b_1}$, $\chi_{b_1}$) are the 
direction of $b_1$ in the $b$ rest frame.
   Since the orientation of $x$-axis is a matter of convention, only the
difference $\chi = \chi_{a_1}-\chi_{b_1}$ of the two azimuthal angles is 
physical.

%======================================================================
%======================================================================

    In most general form the angular distribution for decay 
$B^0(t)\to a(\to a_1a_2)\,b(\to b_1b_2)$ can be expressed as
\begin{equation}
f_0(\Theta_{a_1},\Theta_{b_1}, \chi; t) \equiv
\frac{d^3\Gamma}{d\mbox{cos}\Theta_{a_1} d\mbox{cos}\Theta_{b_1} d\chi}
= \frac{9}{64\pi}\,\sum^6_{i=1} {\cal O}_i(t)
  g_i(\Theta_{a_1},\Theta_{b_1}, \chi)\,,
\label{angle_dist}
\end{equation}
where ${\cal O}_i$ ($i=1,...,6$) are the observables (\ref{observ}) and 
$g_i$ are the functions of physical angles $\Theta_{a_1}$, $\Theta_{b_1}$ 
and $\chi$.

   The time-dependent angular distribution function (\ref{angle_dist}) is 
used as density of the probability function for MC simulation of 
the vertex and kinematics of the final-state particles in case of decay
$B^0(t)\to a(\to a_1a_2)\,b(\to b_1b_2)$.
   The variables of the function $f_0(\Theta_{a_1},\Theta_{b_1},\chi;t)$
can not be factorized and randomly generated in the independent way.
   Nevertheless, the random generation of $\mbox{cos}\,\Theta_{a_1}$, 
$\mbox{cos}\,\Theta_{b_1}$, $\chi$ and $t$ can be performed either 
simultaneously according to distribution function (\ref{angle_dist}) by 
using four-dimensional random generator or successively, one after 
another, by using single-dimensional random number generators with 
accordance to distribution functions obtained by successive integration of
the function (\ref{angle_dist}) over its variables.
   
   Let as consider the latter approach in the case of sequential random
generation of the variables $t$, $\chi$, $\mbox{cos}\,\Theta_{a_1}$ and
$\mbox{cos}\,\Theta_{b_1}$. 
   The following three distribution function are used in this case:
\begin{eqnarray}
&&
f_1(\Theta_{a_1},\chi;t) \equiv \int_{-1}^{+1} d\mbox{cos}\Theta_{b_1}\,
                                  f_0(\Theta_{a_1},\Theta_{b_1},\chi;t)\,,
\nonumber\\ &&
f_2(\chi;t) \equiv \int_{-1}^{+1} d\mbox{cos}\Theta_{a_1}\,
                                  f_1(\Theta_{a_1},\chi;t)\,,
\label{angle_dist_1}
\end{eqnarray}
and
\begin{equation}
f_3(t) \equiv \int_{0}^{2\pi} d\chi f_2(\chi;t)\,
              = {\cal O}_1(t)+{\cal O}_2(t)+{\cal O}_3(t)\,. 
\label{time_dist}
\end{equation}
  The procedure is as follows:
\begin{itemize}
\vspace*{-2mm}
\item First, the proper time $t$ is randomly generated according to the 
      distribution function $f_3(t)$.
\vspace*{-2mm}
\item Second, the angle $\chi$ is randomly generated according to 
      the single-dimensional distribution given by function $f_2(\chi;t)$ 
      with $t$ being fixed to be equal to the time, generated at the first
      step.
\vspace*{-2mm}
\item Then, the value $\mbox{cos}\,\Theta_{a_1}$ is generated according to
      distribution function $f_1(\Theta_{a_1},\chi;t)$ with values of $t$ 
      and $\chi$ fixed to be equal to their values generated at the 
      previous two steps.
\vspace*{-2mm}
\item Finally, the value of $\mbox{cos}\,\Theta_{b_1}$ can be generated 
      according to the single-dimensional distribution given by function 
      $f_0(\Theta_{a_1},\Theta_{b_1},\chi;t)$ with a properly fixed values 
      of $t$, $\chi$ and $\mbox{cos}\,\Theta_{a_1}$.
\end{itemize}
   The functions of Eqs.~(\ref{angle_dist_1}) and (\ref{time_dist}) 
present the most important experimentally observable distributions.

\vspace*{4mm}
%=======================================================================
{\large\bf 5. Decays $B^0(t)\,,\bar{B}^0(t)\to V_1(l^+l^-)V_2(P^+P^-)$}
%=======================================================================
\vspace*{2mm}

   In case of sequential decays 
$B^0(t)\,,\overline{B}^0(t)\to V_1(\to l^+ l^-) V_2(\to P^+ P^-)$,
where $l^\pm$ and $P^\pm$ are lepton and pseudoscalar mesons, respectively,
let associate the particles in the final states with general 
notations used in Fig.~\ref{B_dec_ANGL} as follows:
\vspace*{-4mm}
\begin{eqnarray*}
&& V_1 \to a\,,\quad l^+ \to a_1\,,\quad l^- \to a_2\,,
\\
&& V_2 \to b\,,\quad P \to b_1\,,\quad P^{\prime} \to b_2\,.
\end{eqnarray*}
   In this case the functions $g_i(\Theta_{a_1},\Theta_{b_1}, \chi)$ in 
Eq.~(\ref{angle_dist}) are defined as \cite{kramer}:
\begin{eqnarray}
&& g_1 = 4 \mbox{sin}^2\Theta_{a_1} \mbox{cos}^2\Theta_{b_1}\,,
\nonumber\\
&& g_2 =   (1+\mbox{cos}^2\Theta_{a_1}) \mbox{sin}^2\Theta_{b_1} 
         - \mbox{sin}^2\Theta_{a_1} \mbox{sin}^2\Theta_{b_1} \mbox{cos}2\chi\,,
\nonumber\\
&& g_3 =   (1+\mbox{cos}^2\Theta_{a_1}) \mbox{sin}^2\Theta_{b_1} 
          + \mbox{sin}^2\Theta_{a_1} \mbox{sin}^2\Theta_{b_1} 
            \mbox{cos}2\chi\,,
\nonumber\\
&& g_4 = 2\mbox{sin}^2\Theta_{a_1} \mbox{sin}^2\Theta_{b_1} \mbox{sin}2\chi\,,
\nonumber\\
&& g_5 = -\sqrt{2}\mbox{sin}2\Theta_{a_1} \mbox{sin}2\Theta_{b_1}
                  \mbox{cos}\chi\,,
\nonumber\\
&& g_6 = \sqrt{2}\mbox{sin}2\Theta_{a_1} \mbox{sin}2\Theta_{b_1} 
                 \mbox{sin}\chi\,.
\label{g_VllVpp}
\end{eqnarray}

   The explicit form of the distribution functions (\ref{angle_dist_1}) are 
given below:
\vspace*{-2mm}
\begin{eqnarray}
f_1(\mbox{cos}\Theta_{a_1},\chi;t) &=&
   ~~\frac{4}{3}{\cal O}_1(t)\,\mbox{sin}^2\Theta_{a_1}
  +\frac{4}{3}{\cal O}_2(t)\,(1-\mbox{sin}^2\Theta_{a_1}\mbox{cos}^2\chi)
\nonumber\\ &&
  +\frac{4}{3}{\cal O}_3(t)\,(1-\mbox{sin}^2\Theta_{a_1}\mbox{sin}^2\chi)
  -{\cal O}_4(t)\,\mbox{sin}^2\Theta_{a_1}\mbox{sin} 2\chi\,,
\nonumber\\
f_2(\chi,t) &=&
   ~~{\cal O}_1(t) +\frac{1}{2}{\cal O}_2(t)\,(3-2\mbox{cos}^2\chi)
\nonumber\\ &&
  +\frac{1}{2}{\cal O}_3(t)\,(3-2\mbox{sin}^2\chi)
  -{\cal O}_4(t)\,\mbox{sin} 2\chi\,.
\label{f_12}
\end{eqnarray}

\vspace*{4mm}
%=================================================================
{\large\bf 6. Physics parameters for decay $B^0_s\to J/\psi\phi$}
%=================================================================
\vspace*{2mm}

  For numerical MC simulation of the decay $B^0_s\to J/\psi\phi$ the
following physics parameters should be defined:
\begin{itemize}
\vspace*{-2mm}
\item $\alpha = 1$ -- no CP violation in $B^0-\overline{B}^0$-mixing;
\vspace*{-2mm}
\item $M_L = 5.3696$ GeV -- mass of the ``light'' $B$ meson;
\vspace*{-2mm}
\item $\Delta M = M_H - M_L = $ 10.6 ps$^{-1}$;
\vspace*{-2mm}
\item $\Gamma_L = 1/\tau$ -- width of the ``light'' $B$ meson, where 
      $\tau = 1.493$ ps;
\vspace*{-2mm}
\item $\Delta\Gamma = \Gamma_H - \Gamma_L = 0.2\Gamma_L$;
\vspace*{-2mm}
\item $\phi_j^{(s)} = 0.03\div 0.06$ -- CP-violating weak phase;
\vspace*{-2mm}
\item initial values of observables at time $t=0$: $|A_0(0)|$, 
      $|A_{||}(0)|$, $|A_\bot(0)|$, and two CP-conserving strong phases
      $\delta_1 \equiv \mbox{arg}[A_{||}^*(0) A_\bot(0)]$ and
      $\delta_2 \equiv \mbox{arg}[A_0^*(0)A_\bot(0)]$.
\end{itemize}
\vspace*{-2mm}

  The time-reversal invariance and naive factorization lead to the
following common property:
$$
\mbox{Im}[A_0^*(0)A_\bot(0)]     = 0\,,\quad
\mbox{Im}[ A_{||}^*(0)A_\bot(0)] = 0\,,
$$
$$
\mbox{Re}[A_0^*(0)A_{||}(0)]     = \pm |A_0(0) A_{||}(0)|\,.
$$
   Moreover, in absence of strong final-state interactions, 
$\delta_1= \pi$ and $\delta_2=0$.

   Using low-energy effective Hamiltonian \cite{fleischer2} for nonleptonic 
$b$-quark transitions $b\to s\bar{c}c$, the amplitudes $A_f(0)$ 
($f = 0,||,\bot$) of $B^0_s\to J/\psi\,\phi$ decay can be calculated within
the framework of naive factorization in terms of the form factors 
of transitions $B\to\phi$ induced by quark currents.
   The $B\to\phi$ form factors can be related to the $B\to K^*$ case by
using SU(3) flavor symmetry.
   In table~1 we collect the predictions of Ref.~\cite{dighe2} for the ratios
of $B^0_s\to J/\psi\,\phi$ observables calculated with $B\to K^*$ form 
factors given by different models \cite{BSW}--\cite{cheng}.
   The quantity
$$
\frac{\Gamma_0(0)}{\Gamma_0(0) + \Gamma_T(0)} =
\frac{|A_0(0)|^2}{|A_0(0)|^2 + |A_{||}(0)|^2 + |A_\bot(0)|^2}
$$
describes the ratio of the longitudinal to the total rate at $t=0$.

%=========================================================================
\vspace*{2mm}
\begin{center}
Table~1. Predictions for $B^0_s\to J/\psi\,\phi$ observables 
\end{center}
\def\largelinestretch{\renewcommand{\baselinestretch}{1.5}}
    \largelinestretch\normalsize
\vspace*{-4mm}
\begin{center}
\begin{tabular}{|c|c|c|c|}
\hline
Observable & BSW \cite{BSW}& Soares \cite{soares}& Cheng \cite{cheng}\\
\hline
\hline
$|A_{||}(0)|/|A_0(0)|$      & 0.81 (0.77) & 0.82 (0.78) & 0.75 (0.70) 
\\ \hline
$|A_\bot (0)|/|A_0(0)|$     & 0.41 (0.40) & 0.89 (0.88) & 0.55 (0.54) 
\\ \hline
$\Gamma_0(0)/(\Gamma_0(0) + \Gamma_T(0))$
                            & 0.55 (0.57) & 0.40 (0.42) & 0.54 (0.56)
\\\hline
\end{tabular}
\end{center}
\def\largelinestretch{\renewcommand{\baselinestretch}{1.1}}
\largelinestretch\normalsize
%=========================================================================

\vspace*{4mm}
%===================================================
{\large\bf 7. General structure of program SIMUB}
%===================================================
\vspace*{2mm}

   The package for generation of production of $B$-mesons and their decays,
SIMUB, is kept under the directory {\bf \tt SIMUB\_V\_X} 
({\tt X} is the version number) which has the following main parts:
\begin{itemize}
\vspace*{-2mm}
\item {\tt bb\_gen} -- routines needed to generate $b\bar{b}$ events
                       (FORTRAN, PYTHIA, HBOOK);
\vspace*{-2mm}
\item {\tt bb\_frg} -- routines performing string fragmentation and generation
                       of $B$-mesons (FORTRAN, PYTHIA, HBOOK).
                       The program is a part of {\tt BB\_dec} program but may 
                       be used as independent program;
\vspace*{-2mm}
\item {\tt BB\_dec} -- routines performing $B$-decays 
                       (C++, FORTRAN, PYTHIA, HBOOK, ROOT) and
                       storing the results into standard HEPEVT-format
                       Ntuple for further usage in the CMS detector
                       simulation or in format JETSET for analysis;
\vspace*{-2mm}
\item {\tt include} -- a collection of common blocks for {\tt bb\_gen},
                       {\tt bb\_frg} and {\tt BB\_dec}.
\vspace*{-2mm}
\end{itemize}
   To simulate the $B$-meson production in $pp$ collisions with PYTHIA
\cite{PYTHIA} (programs {\tt bb\_gen} and {\tt bb\_frg}) we used the modified 
routines from the FORTRAN-based package described in \cite{konecki}.

   The data flow between three main parts ({\tt bb\_gen}, {\tt bb\_frg},
and {\tt BB\_dec}) of the package {\tt SIMUB\_V\_X} is organized in the 
following two steps:
\begin{itemize}
\vspace*{-2mm}
\item At the first step the $b\bar{b}$-events are generated by the program
      {\tt bb\_gen} and written to Ntuple {\tt bb\_X\_YYY}.
\vspace*{-2mm}
\item At the next step the program {\tt BB\_dec} reads $b\bar{b}$-events from
      Ntuple {\tt bb\_X\_YYY} and performs the string fragmentation (via call
      {\tt bb\_frg}) and decay of $B$ mesons. The decay modes and mechanisms
      are defined by user. The full information about $B$ decay events is 
      stored in Ntuple {\tt BB\_dec.ntpl}. Its format (HEPEVT or JETSET) is 
      defined by user.
\end{itemize}

\vspace*{4mm}
%==================================================================
{\large\bf 8. Inheritance of classes in the program {\tt BB\_dec}}
%==================================================================
\vspace*{2mm}

\begin{figure}[hbt]
\begin{center}
\includegraphics[width=0.8\textwidth] {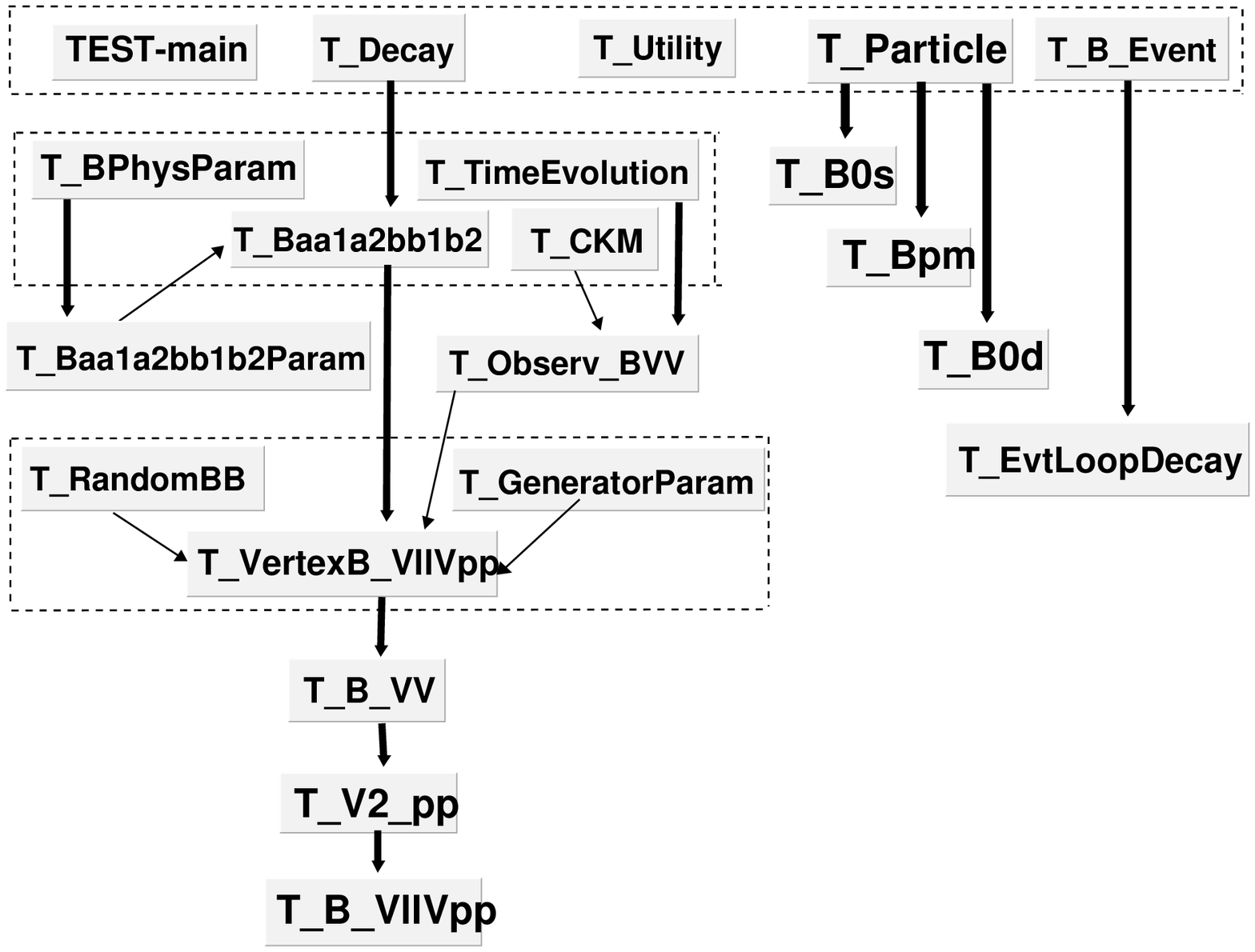}
\end{center}
\vspace*{-10mm}
\caption{\small {The class inheritance of the {\tt BB\_dec} program
                 in case of simulation of decays
  $B^0(t)\,,\overline{B}^0(t)\to  V_1(\to  \l^+\l^-) V_2(\to P^+P^-)$
                 }
         }
\label{BB-dec-struct}
\end{figure}

   The example of class inheritance of {\tt BB\_dec} program is shown in
Fig.~\ref{BB-dec-struct} for the case of time-depended decay channel
\vspace*{-2mm}
$$
  B^0(t)\,,\overline{B}^0(t)\to V_1 (\to  l^+l^-)\, V_2 (\to  P^+P^-)\,,
\vspace*{-2mm}
$$
where $V_1$ and $V_2$  are the vector particles,
$l^+$ and $l^-$ are the leptons,
$P^+$ and $P^-$ are the pseudoscalar mesons.
   The thick and thin arrows show the directions of class inheritance from 
mother to daughter classes. 
   Thick arrows correspond to main inheritance ways.
   The name of each class coincides with the name of subdirectory where this 
class is kept.
   The upper dashed-line box corresponds to the directory {\tt src}. 
   The lower dashed-line boxes unite the names of classes which are in the same
subdirectory at the same level.
   In case when several classes belong to the same subdirectory, its name
coincides with the name of the class where the thick incoming arrow shows to.
   The directories are included one into another in the directions opposite 
to thick arrows.

%==========================================================================
%==========================================================================

   The main directory {\tt src}, shown in Fig.~\ref{BB-dec-struct} as upper
dashed-line box,  contains five subdirectories {\tt T\_Decay}, 
{\tt T\_Particle}, {\tt T\_B\_Event}, {\tt T\_Utility} and {\tt TEST-main}.
   The subdirectory {\tt T\_Decay} contains definition of the mother class 
with the same name as well as subdirectories with classes which are 
common for different decay channels.
   The subdirectory {\tt T\_Particle} contains the definition of the mother
class with the same name and subdirectories with daughter classes 
{\tt T\_B0s}, {\tt T\_Bpm} and {\tt T\_B0d} which define the properties and 
decay modes for $B^0_d$ ($\overline{B}^0_d$), $B^{\pm}$ and $B^0_s$ 
($\overline{B}^0_s$) mesons, respectively. 
   The directory {\tt T\_B\_Event} contains the definition of the mother class
with the same name and daughter class {\tt T\_EvtLoopBDec} with a loop over 
all events with $B$-mesons.
   In the loop, each event is read from the ROOT Tree {\tt bb\_frg.root} and, 
then, the decays of $B$-mesons are performed according to modes defined in the 
samples of classes {\tt T\_B0s}, {\tt T\_Bpm} and {\tt T\_B0d} both for 
particles and anti-particles.
   The directory {\tt T\_Utility} consists of the auxiliary classes and 
functions.
   The directory {\tt TEST-main} contains the main programs for testing of 
classes.

   The directory {\tt src} also includes the file {\tt T\_EvtLoopBdec\_main.C}
containing a {\tt main()} function which is not shown in 
Fig.~\ref{BB-dec-struct}.
   In this file, user can find the example of loop over the events.

\vspace*{4mm}
%==================================================================
{\large\bf 9. Monte Carlo methods for generation of decays 
\hspace*{14mm}$B^0(t)\to V_1(\to\l^+\l^-)\,V_2(\to P^+P^-)$
}
%==================================================================
\vspace*{2mm}

  In class {\tt T\_VertexB\_VllVpp}, the variables describing the 
decay channel and MC method for random generation are defined.
  The member function {\tt T\_VertexB\_VllVpp::DecayB\_VllVpp} generates four
random numbers -- proper decay time $t$ and angular variables 
$\mbox{cos}\,\Theta_{b_1}$, $\mbox{cos}\,\Theta_{a_1}$ and $\chi$ -- according
to distribution function given by Eq.~(\ref{angle_dist}).
   Then, {\tt DecayB\_VllVpp} calculates Lorenz vector characterizing the
decay vertex in Laboratory system.

   Two MC methods of random numbers generation are implemented in 
the class {\tt T\_VertexB\_VllVpp}.

   The first method is based on the filling of the large single-dimensional 
array {\tt f4\_Integ[n4\_cells]} of real numbers which represents numerically
the four-dimensional distribution function corresponding to  
$f_0(\Theta_{a_1},\Theta_{b_1}, \chi; t)$ given by Eqs.~(\ref{angle_dist})
and (\ref{g_VllVpp}).
   The array {\tt f4\_Integ} contains 
${\tt n4\_cells}={\tt fNpx}\times{\tt fNpy}\times{\tt fNpz}\times {\tt fNpt}$
elements, where the variables {\tt fNpx}, {\tt fNpy}, {\tt fNpz} and {\tt fNpt}
define the MC generator resolutions.
   The array {\tt f4\_Integ} is filled in the constructor of class 
{\tt T\_VertexB\_VllVpp} where the memory space for the array is reserved 
during all time of the existence of this class sample.
   The array {\tt f4\_Integ} is used in the member function 
{\tt T\_VertexB\_VllVpp::GetRandom4} performing fast generation of the four 
random values of $\mbox{cos}\,\Theta_{b_1}$, $\mbox{cos}\,\Theta_{a_1}$,
$\chi$ and $t$
\footnote{In the function {\tt GetRandom4} we used the algorithm which is 
          analogous to that was realized in the ROOT class {\tt TF3} 
          \cite{ROOT} for generation of three random numbers distributed 
          according to 3-dimensional probability function.}.

   The second method involves sequential generation of random numbers 
according to approach which is based on usage of single-dimensional 
distribution functions given by Eqs.~(\ref{time_dist}) and (\ref{f_12}).
   This method was already described in section 4.
   In this case the constructor {\tt T\_VertexB\_VllVpp} does not fill any
arrays and, therefore, it does not reserve the memory and may be 
used for definition of big number of particles decaying in the same mode.

   These methods have been tested in case of MC simulation of the ``golden'' 
decay mode 
$B^0_s(t),\overline{B}^0_s(t)\to J/\psi(\to\mu^+\mu^-)\,\phi(\to K^+K^-)$
with values of physics parameters fixed according to section 6.
   Both methods demonstrated the identical results with suitable time and 
memory consuming.

   The time-dependent angular correlations in the decays of type
$B^0(t)\to V_1(\to\l^+\l^-)\,V_2(\to P^+P^-)$ can be used in data analysis to 
extract CP-violating asymmetries, $B^0-\overline{B}^0$ mixing parameters 
and other observables \cite{dighe2,valencia,dunietz}.
   Therefore, the decay dynamics described by angular correlations 
(\ref{angle_dist}) should be included without fail into MC generators 
developed for physics studies.

\vspace*{-8mm}
%==========================================================================
\begin{figure}[hbt]
\begin{center}
\includegraphics[width=0.85\textwidth] {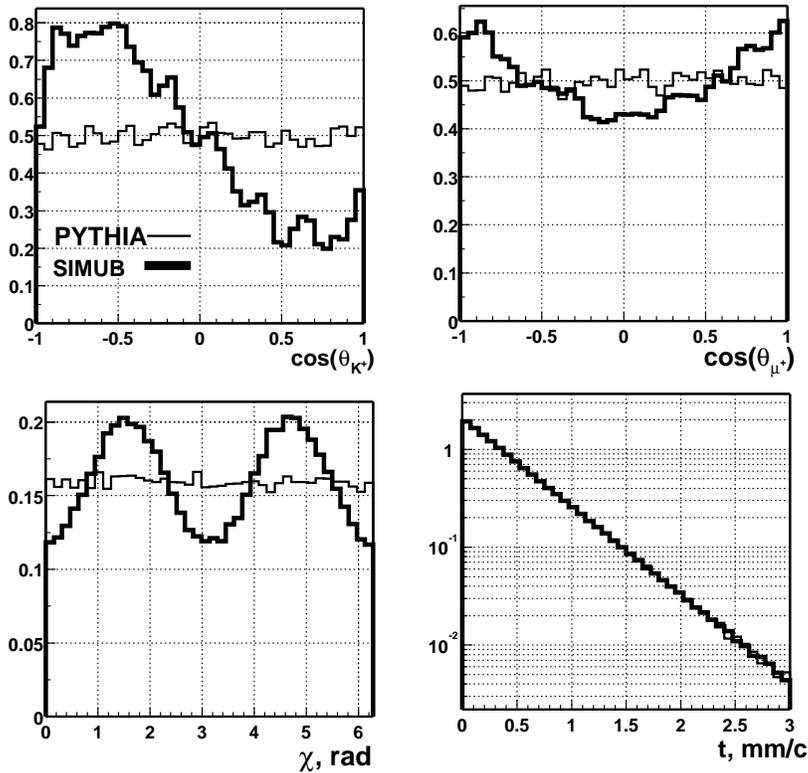}
\end{center}
\vspace*{-10mm}
\caption{\small{Comparison of time and angular distributions for decay 
                $B^0_s(t),\overline{B}^0_s(t)\to J/\psi(\to \mu^+\mu^-)\,
                \phi(\to K^+K^-)$ generated by SIMUB and PYTHIA. For time
                $t$ we use units 1mm/$c\approx 3.33\times 10^{-12}$ sec
               }
        }
\label{Slices}
\end{figure}
%==========================================================================

   In Fig.~\ref{Slices} we compare time and angular distributions in the
helicity frame for decay 
$B^0_s(t),\overline{B}^0_s(t)\to J/\psi(\to\mu^+\mu^-)\,\phi(\to K^+K^-)$ 
generated by SIMUB and PYTHIA.
   The difference between these two generators becomes more clear when slices 
over time and angular variables are cut from the kinematical phase space.
   The angular distributions shown in Fig.~\ref{Slices} are built for
the following kinematical slices:
\vspace*{-2mm}
\begin{eqnarray*}
\chi:                      && 0.01~\mbox{mm/}c<t<0.19~\mbox{mm/c}\,;\\
\mbox{cos}\,\Theta_{\mu^+}:&& 0.01~\mbox{mm/}c<t<0.19~\mbox{mm/c}\,,\quad
                              0.3~\mbox{rad}<\chi< 0.6~\mbox{rad}\,;\\
\mbox{cos}\,\Theta_{K^+}:  && 0.01~\mbox{mm/}c<t<0.19~\mbox{mm/c}\,,\quad
                              0.3~\mbox{rad}<\chi <0.6~\mbox{rad}\,,\\
&&                            0.7<\mbox{cos}\,\Theta_{\mu^+}<1\,.
\vspace*{-4mm}
\end{eqnarray*}

    In case of PYTHIA usage, Fig.~\ref{Slices} shows the uniform distributions
for angular variables $\mbox{cos}\,\Theta_{b_1}$, $\mbox{cos}\,\Theta_{a_1}$ 
and $\chi$ because of lack of time-dependent angular correlations.
   Due to this simple reason PYTHIA can not be used for MC studies of dynamics
of sequential two body decays of $B$ mesons in the channels with 
intermediate vector mesons.
   Unfortunately, due to various technical reasons, the both other well known 
packages, QQ~\cite{QQ} and EvtGen~\cite{EvtGen}, turn out to be also not 
suitable for study of angular correlations in decays of the type
$B\to V_1(\mu^+\mu^-)\, V_2(P^+P^-)$.

\vspace*{4mm}
%======================
 {\large\bf Conclusion}
%======================
\vspace*{2mm}

   General scheme of program chain and data flow for MC simulation of
$B$-meson production and decays have been developed and realized in the 
package SIMUB which is now in progress.
   This scheme was tested for case of semileptonic decays $B\to D^{(*)}\mu\nu$
and sequential decays of the type $B\to V_1(\to\mu^+\mu^-)\,V_2(\to P^+P^-)$.
   The SIMUB package provides unique tools for MC physics studies of dynamics
of semileptonic and "golden" $B$-decay channels with taking into account
$B^0-\overline{B}^0$ oscillations and angular correlations.

   The starting version of the package {\tt SIMUB\_V\_0} is installed and 
tested at CERN on Linux platform.
   The package and documentation for user one can find on the SIMUB Home Web
Page \cite{SIMUB}.
   There is link to this page from CMS $B$-Physics Group Web Page:
{\tt http://cmsdoc.cern.ch/\~\,bphys}. 
   Now the SIMUB is used within CMS collaboration for exclusive $B$-trigger 
studies.

%==========================================================================

\end{document}